\pdfoutput=1
\documentclass[floatfix,reprint,aps,prb,twocolumn,superscriptaddress]{revtex4-1}
\usepackage{graphicx}
\usepackage{dcolumn}
\usepackage{bm}
\usepackage{amsmath}
\usepackage[english]{babel}
\usepackage[utf8]{inputenc}
\usepackage{color}
\begin{document}

\title{Multiplexed quantum transport using commercial off-the-shelf CMOS at sub-kelvin temperatures }

\author{B. Paquelet Wuetz} 
\thanks{These two authors contributed equally}
\author{P. L. Bavdaz}
\thanks{These two authors contributed equally}
\author{L. A. Yeoh}
\affiliation{QuTech and Kavli Institute of Nanoscience, Delft University of Technology, PO Box 5046, 2600 GA Delft, The Netherlands}
\author{R. Schouten}
\author{H. van der Does}
\author{M. Tiggelman}
\affiliation{QuTech, Delft University of Technology, PO Box 5046, 2600 GA Delft, The Netherlands}
\author{D. Sabbagh}
\affiliation{QuTech and Kavli Institute of Nanoscience, Delft University of Technology, PO Box 5046, 2600 GA Delft, The Netherlands}
\author{A. Sammak} 
\affiliation{QuTech and TNO, Stieltjesweg 1, 2628 CK Delft, The Netherlands}
\author{C. G. Almudever} 
\affiliation{QuTech and Computer Architecture Lab, TU Delft, P.O. Box 5046, 2600 GA Delft, The Netherlands}
\author{F. Sebastiano} 
\affiliation{QuTech and Kavli Institute of Nanoscience, Delft University of Technology, PO Box 5046, 2600 GA Delft, The Netherlands}
\author{J.S. Clarke} 
\affiliation{Components Research, Intel Corporation, 2501 NW 229th Ave, Hillsboro, OR 97124, USA}
\author{M. Veldhorst}
\author{G. Scappucci}
\email{g.scappucci@tudelft.nl}
\affiliation{QuTech and Kavli Institute of Nanoscience, Delft University of Technology, PO Box 5046, 2600 GA Delft, The Netherlands}
\date{\today}

\begin{abstract}
Continuing advancements in quantum information processing have caused a paradigm shift from research mainly focused on testing the reality of quantum mechanics to engineering qubit devices with numbers required for practical quantum computation. One of the major challenges in scaling toward large-scale solid-state systems is the limited input/output (I/O) connectors present in cryostats operating at sub-kelvin temperatures required to execute quantum logic with high-fidelity. This interconnect bottleneck is equally present in the device fabrication-measurement cycle, which requires high-throughput and cryogenic characterization to develop quantum processors. Here we multiplex quantum transport of two-dimensional electron gases at sub-kelvin temperatures. We use commercial off-the-shelf CMOS multiplexers to achieve an order of magnitude increase in the number of wires. Exploiting this technology we advance 300 mm epitaxial wafers manufactured in an industrial CMOS fab to a record electron mobility of (3.9$\pm$0.6)$\times$10$^5$ cm$^2$\slash Vs and percolation density of (6.9$\pm$0.4)$\times$10$^{10}$ cm$^{-2}$, representing a key step toward large silicon qubit arrays. We envision that the demonstration will inspire the development of cryogenic electronics for quantum information and because of the simplicity of assembly, low-cost, yet versatility, we foresee widespread use of similar cryo-CMOS circuits for high-throughput quantum measurements and control of quantum engineered systems.    
\end{abstract}

\maketitle

\section{Introduction}

With quantum computing technology advancing at a fast pace, noisy intermediate-scale quantum (NISQ) technology with 50-100 qubits are predicted to be realized in the near future.\cite{preskill2018quantum, terhal2018quantum} Solid-state quantum processors in the NISQ era and beyond will be realized by mass-fabrication on wafers including 300 mm technology.\cite{Clarke2016,pillarisetty2019qubit,SabbaghPhysRevApplied2019} Optimization and validation approaches for quantum materials and devices are therefore required that can rely on an increasingly fast-feedback cycle. Since quantum technology operates at sub-kelvin temperatures, cryogenic solutions for fast testing will have to be developed.

The decades of advancement in classical technology following Moore's law has been made possible by approaches dictated by Rent's rule  $T=tg^p$, where the Rent exponent $P$ relates the total number of control lines $T$ and proportionality factor $t$ with the number of internal components $g$.\cite{landman1971pin,lanzerotti2005microminiature} This same rule has been predicted to be required for practical quantum processors,\cite{franke2019rent} but we also envision that this rule will determine the progress in fabrication and validation, with the Rent factor crucially determining how many devices can be tested simultaneously.

One pursuit toward scalable testing is to adapt room temperature wafer-scale probing at cryogenic temperatures. Indeed, a cryogenic wafer prober has recently been developed to establish a high-volume 300 mm test-line for quantum devices.\footnote{{https://newsroom.intel.com/news/intel-drives-development-quantum-cryoprober}, accessed: 2019-07-01} The measurement temperature in probe-based systems, however, is limited to a few Kelvin. Furthermore, integration of magnets required for material characterization is challenging to achieve on large-size probe systems. Alternatively, cryogenic on-chip multiplexers have been developed in GaAs/AlGaAs\cite{MUXCMBR1,MUXCMBR2,MUXCMBR3,MUXCMBR4} and Si/SiGe\cite{MUXEriksson} heterostructures, operating at a temperature of 1.6 K and 0.2 K, respectively. With this approach the number of quantum devices measured in one cooldown on a single chip is increased without the need to alter existing cryostat setups. However, the design and implementation of on-chip multiplexers is specific to the materials and device under test (DUT). Furthermore, an architecture that works at base temperature of a dilution refrigerator, high magnetic fields, and is independent of the number and type of DUT has yet to be developed.

\begin{figure}
	\includegraphics[width=85mm]{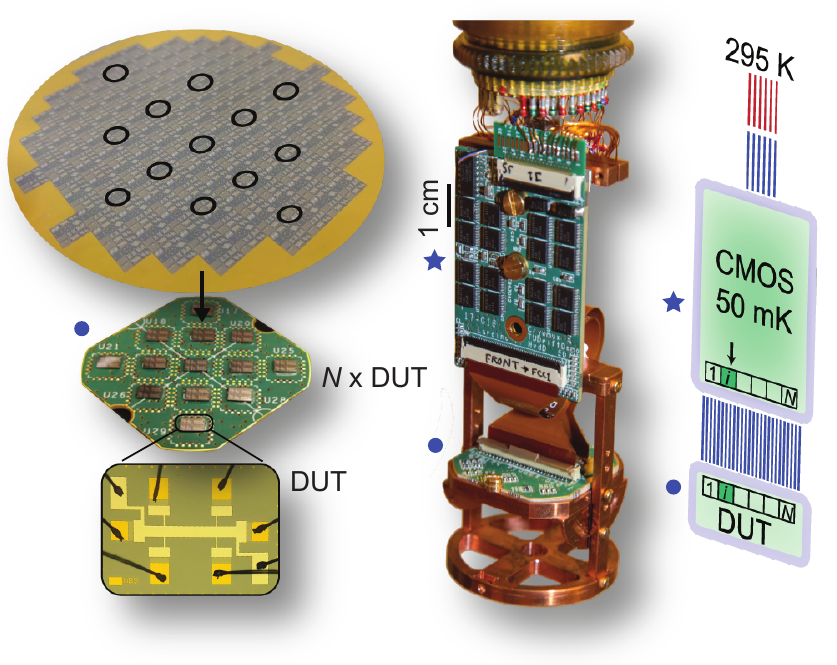}%
	\caption{Setup for accelerated testing and validation of quantum materials and devices  using CMOS at sub-kelvin temperatures. Left panel: a number $N$ of dies, each containing a device under test (DUT) are selected from a wafer and wire bonded onto a printed circuit board (DUT-PCB; blue circle). Middle panel: the DUT-PCB is mounted to the cold finger of a dilution refrigerator (MCK 50-400 by Leiden Cryogenics) connected by flat ribbon cables to a printed circuit board containing CMOS components (cryo-MUX PCB; blue star). The DUT-PCB and cryo-MUX PCB are operated at $T$ = 50 mK. Right panel: schematics of the cold finger showing how the use of cryo-CMOS allows cold-wires multiplication on the DUT-PCB with a fixed overhead of wires to room temperature. Devices may be selected for single measurements or time division multiplexing} 
\label{fig:mux}
\end{figure}

Here we deploy digital CMOS logic at $T$ = 50 mK to increase the number of wires available at cryogenic temperature by an order of magnitude while keeping the overhead number of I/O wires at room temperature fixed (Fig. \ref{fig:mux}). Our cryogenic platform is based on low-cost commercial off-the-shelf multiplexers driven by a nearby shift-register, is operated under the extreme temperature and magnetic fields achieved in dilution refrigerators, and can be readily integrated in any kind of cryostat. We have specifically designed the cryo-CMOS circuit to act as a switch and allow for high-throughput quantum transport measurements. Multiple devices can be screened for relevant metrics in the same cooldown either individually or at once by time-division multiplexing (TDM).  

The paper is organized as following: we first describe the cryo-CMOS system design and elaborate on the scaling properties of the architecture. We demonstrate that using commercial off-the-shelf components no artifacts are introduced in the multiplexed measurement of calibration resistors while sweeping parameters such as voltage bias, magnetic field, and temperature. To prove the value of this architecture for accelerating the fabrication-measurement cycle of quantum devices, we focus on an archetypal measurement in condensed matter physics: magnetotransport of 2DEGs in the classical and quantum Hall regime. These measurements are used to evaluate statistically key metrics of high-mobility Si/SiGe heterostructures field effect transistors, relevant for spin-qubits in Si,\cite{zajac_resonantly_2018,watson_programmable_2018} currently leading the field of quantum computation with quantum dots.\cite{Loss1998} We exploit the cryo-multiplexing platform to advance 300 mm epitaxial wafers manufactured in an industrial CMOS fab to record values of electron mobility and percolation density at sub-kelvin temperatures, relevant for large silicon spin-qubit arrays.

\section{Results}
\subsection{A cryogenic multiplexer platform}

\begin{figure}%
	\includegraphics[width=85mm]{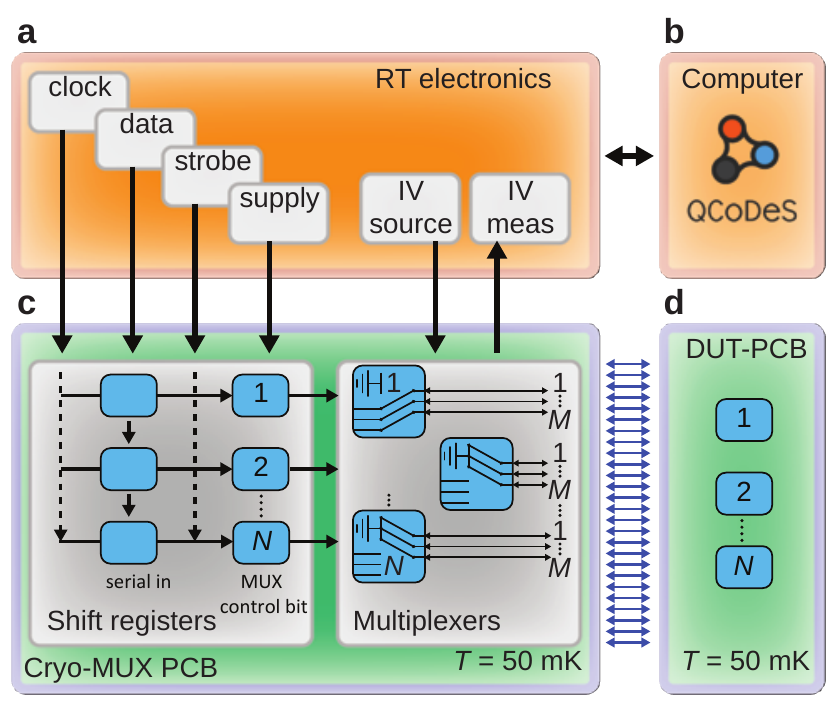}%
	\caption{Cryo-multiplexing platform setup. \textbf{a} Electronics operated at room temperature is controlled by a computer \textbf{b} using the QCoDeS framework and supplies voltages to the components located on the cryo-MUX PCB \textbf{c}. The serial-input parallel-output shift registers receive a string of bits from the room temperature electronics to control the multiplexers. Each bit corresponds to all multiplexers associated with one device under test, located on the DUT-PCB \textbf{d}. The multiplexed lines of select devices can be switched either to the supply and measurement equipment or to ground.} 
\label{fig:cryomux}
\end{figure}

Figure \ref{fig:cryomux} shows schematics of our experimental setup. At the heart of the architecture is a printed circuit board (cryo-MUX PCB; Fig. \ref{fig:cryomux}c) operating at 50 mK. The cryo-MUX PCB comprises cascaded serial-input parallel-output (SIPO) shift registers which provide $N$ outputs lines, each of them controlling $M$ outputs lines of multiplexer components. Few input/output (I/O) wires connect the cryo-MUX PCB to room temperature electronics (Fig. \ref{fig:cryomux}a) with the following purpose: i) provide supply voltages and digital logic levels to the board components; ii) connect the multiplexers to current/voltage supplies and equipment for performing measurements of the devices. Each of them has $M$ ($S$) multiplexed (shared) terminals and are bonded to a printed circuit board (DUT-PCB; Fig. \ref{fig:cryomux}d). The DUT-PCB, also operating at 50 mK, is connected to the multiplexers on the cryo-MUX PCB by flat ribbon I/O cables supporting more than $NM+S$ wires.

Table \ref{tab:lines_scaling} presents an overview of the scaling properties of the number of lines between the different parts of the cryogenic architecture in our experimental setup. The system can be scaled by either adding more devices or by adding more lines per device, i.e increasing $N$ or $M$, respectively. When devices are added, additional shift registers are required to select these devices. When the number of lines per device is increased, additional multiplexers components and room temperature measurement equipment are needed. Crucially, this protocol requires a constant number of lines between room temperature and cryogenic components, regardless how large $N$ is. This approach yields an optimal Rent exponent at room temperature $p_{RT}$=0, however the time necessary to perform a measurement cycle through all DUT scales linearly with $N$.

\begin{table}
\begin{tabular}{l|llll}
                & RT & Shift & Mutliplexers & DUT   \\ 
                & electronics & registers & & \\ \hline
RT electronics  &              &C$_1$            & M+C$_2$      &  S    \\
Shift registers & C$_1$          &               & N            &       \\
Mutliplexers    & M+C$_2$        & N               &             & NM    \\
DUT             & S              &                 & NM           & 
\end{tabular}
\caption{Number of lines between parts of the cryogenic multiplexer platform. A constant number is indicated with $C$, whereas $N$ is the number of DUT and $M$ ($S$) is the number of multiplexed (shared) lines per DUT. The first row indicates that the lines between room temperature and cryogenic temperatures are not dependent on $N$. On the other hand, scaling the system  will increase the lines between the multiplexers and DUT.} 
\label{tab:lines_scaling}
\end{table}

The whole system is controlled by sending commands to the electronics through a software environment built on QCoDeS\footnote{https://github.com/QCoDeS/Qcodes, accessed: 2019-07-01} (Fig. \ref{fig:cryomux}b), while timing is done using an internal hardware clock for increased precision. Three signals generated from custom digital to analog converters are sent to the SIPO shift registers to perform switching between DUT. All signals are produced by low-noise equipment to avoid any coupling of noise and interference to the multiplexers, since no specific shielding/protection could have been adopted at sub-kelvin temperatures. Firstly, a sequence of data bits is sent that defines which DUT will be selected. Secondly, a clocking signal is sent while loading each bit. Thirdly, a strobe signal is supplied, indicating when the shift register is fully loaded and the outputs can be sent to the multiplexers. 

To achieve switching between lines, each line in the DUT is connected to a multiplexer consisting of a CMOS analog integrated circuit configured as a single-pole/double-throw switch. The input terminal of the switch is connected to the DUT, while the output terminals are connected to room temperature equipment and ground. All switches associated with a DUT are controlled through logic inputs connected to the same shift register output. All possible 2$^N$ combinations of DUT can be selected since the multiplexers are driven by the parallel output of the shift register. 

In all the experiments presented here, the cryo-MUX PCB comprises two cascaded shift registers with eight parallel output each (Texas Instruments 74HC4094; specifications in Ref. \footnote{https://www.ti.com/lit/ds/symlink/cd74hc4094.pdf, accessed: 2019-07-01}), allowing, in principle to measure up to $N$=16  DUT. Each of the $N$ parallel outputs of the shift registers control $M$=6 multiplexed lines, separated over two three single-pole/double-throw switches components (Maxim  MAX4619; specifications in Ref. \footnote{https://datasheets.maximintegrated.com/en/ds/MAX4617-MAX4619.pdf, accessed: 2019-07-01}. These components show an on-resistance of 7 $\Omega$ and an off-resistance $\geq$ 2 G$\Omega$ at cryogenic temperatures, limited by the measurement setup. The shift registers and multiplexer components are powered with positive and negative supply voltages of 1.1 V and -3.9 V, respectively. The same values define the digital logic levels. In total, the 16 available channels and 6 multiplexed lines result in 96 wires available at the base temperature of the dilution refrigerator. Up to 13 devices are bonded on the the DUT-PCB, less than $N$=16 due to the specific die-size chosen for the DUT and the limited sample space. A complete circuit diagram of the cryo-MUX PCB and DUT-PCB is provided in the Supplementary Information.

We are able to discriminate whether correct switching has occurred by monitoring the resistance of $N$ control resistors, each connected to one of the $M$ = 6 multiplexed lines. The switching success rate, i.e. the ability to successfully select a desired DUT for measurement, is determined by sending commands to switch between random control resistors and after many switches (up to 10$^5$) by comparing the measured resistance to the expected resistance. The switch is considered successful when these values match. We obtain a switching success rate of 100\% at 250 mK, with a clock frequency of 4.4 MHz and a strobe frequency of $\approx$ 8 KHz. 

While switching, the transistors in the multiplexers dissipate heat as a function of switching frequency $f$, which is set to 8 Hz in the experiments reported below. We estimate the dissipation caused by switching one component at 8 Hz to be $\approx$1.5 nW, well below the 400 $\mu$W cooling power of our dilution refrigerator at 100 mK. This dissipation of the multiplexer is extrapolated from the linearly increasing dissipation with frequency of 0.19 nW/Hz measured up to $f\geq$100 kHz at cryogenic temperatures.   

\subsection{Time-division multiplexing of known resistors upon bias, magnetic field, and temperature sweeps}

\begin{figure}
	\includegraphics[width=85mm]{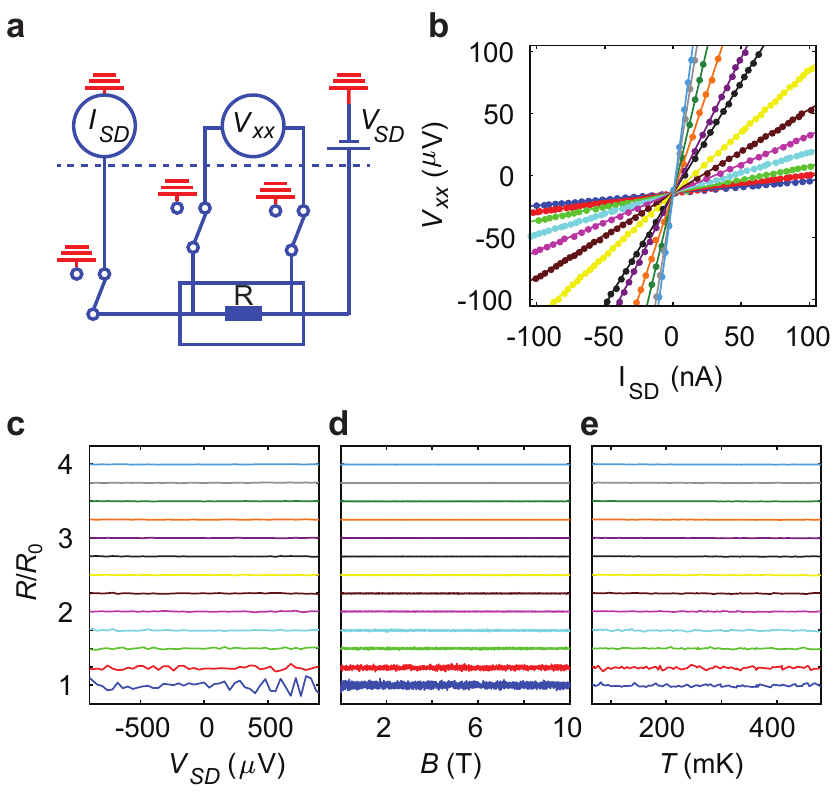}%
	\caption{\textbf{a} Four-probe setup for multiplexed measurements of known resistors. \textbf{b} Dc voltage-current characteristics of thirteen resistors measured individually (dots) or all at once (lines) by time division multiplexing. Different colors correspond to different resistors. \textbf{c}-\textbf{e} Multiplexed resistance measurements while sweeping source-drain voltage ($V_{SD}$), magnetic field ($B$), and temperature ($T$). On the vertical axis, the AC resistance $R$ = $dV_{xx}/dI_{SD}$ is normalized to the resistance value $R_0$ measured at zero dc source-drain voltage, zero magnetic field, and $T$ = 50 mK. For clarity, curves are offset by an amount 0.25$j$, with $j$ being an integer from 0 (bottom curve) to 12 (top curve).} 
\label{fig:known_resistors}
\end{figure}

Thirteen metal thin film resistors ($N$=13) are bonded to the DUT-PCB in a four-probe configuration (Fig. \ref{fig:known_resistors}a) to validate multiplexed electrical transport under different control sequences and conditions of external parameters, such as source-drain voltage applied to the resistors ($V_{SD}$), magnetic field ($B$), and temperature ($T$). The four-probe setup eliminates the series resistance originating from fridge wiring and electrical contacts and is a testbed for quantum devices characterization. At room temperature the resistance of the chosen components ranges from 100 $\Omega$ to 8.2 $k\Omega$ and is expected to be temperature independent, minimizing device unpredictability. 

We investigate two measurements protocols. Firstly, the cryo-MUX PCB may act as a simple DUT-selector by keeping a single device connected to the measurement equipment whilst sweeping the relevant parameter. This allows for a traditional single device measurement, with $N$ devices eventually measured one after the other. Alternatively, TDM is achieved by sequentially selecting for measurement all resistors at each point in the parameter sweep, allowing all $N$ measurements to be completed within a single parameter sweep. In addition to benefiting from measurement speedup, this protocol allows for an easy comparison between devices since differences in time-dependent factors are minimized. 

In Fig. \ref{fig:known_resistors}b we compare the dc voltage-current characteristics of the resistors obtained by sweeping the source-drain voltage $V_{SD}$ following the two methodologies (sequential sweeps vs TDM). For all resistors the curves obtained with the two methodologies are matching, with fitted resistance values differing only by 0.7\%. Having established the validity of the TDM methodology, we further test its applicability to $V_{SD}$, $B$, or $T$ sweeps, to emulate typical quantum transport measurements. For these measurements we use four-terminal low-frequency lock-in techniques by applying constant AC source-drain voltages of 100 $\mu$V. As seen in Fig. \ref{fig:known_resistors}c-e, the resistance values remain constant for all $N$ devices while sweeping $V_{SD}$, $B$, or $T$. Overall, this characterisation indicates that TDM does not introduce non-linearity in the four terminal measurements and that the whole architecture works properly under high magnetic fields and different temperature conditions. 

\subsection{Multiplexed quantum transport of industrial Si/SiGe field effect transistors}

We now harness the power of the multiplexing platform to measure quantum transport of buried-channel semiconductor heterostructures, an archetype material platform for the fabrication of gated semiconductor quantum devices. In Si/SiGe heterostructures a type II band alignment promotes electron confinement at the interface between a strained Si quantum well and a SiGe barrier.\cite{schaffler1997high} Si/SiGe heterostructures fabricated in academic environments have proven a successful material platform for obtaining long-lived high-fidelity electron spin-qubits in silicon.\cite{yoneda2018quantum} Furthermore, the advanced level of quantum control in these qubits allows to run quantum algorithms on two qubit processors.\cite{zajac_resonantly_2018,watson_programmable_2018}  

By investigating quantum transport in Hall-bar shaped heterostructures field effect transistors,\cite{laroche2015scattering,mi2015magnetotransport,lu2007capacitively} key material metrics such as maximum mobility and percolation density are extracted. Electron mobility is a straightforward figure of merit to asses the overall quality of the 2DEG in the high  density regime, where screening of impurity scattering is relevant.\cite{kim2017annealing,tracy2009observation} On the other hand, the percolation density indicates the minimum density necessary to establish a metallic conduction channel and is a gauge for disorder at low density, where quantum devices operate. 

In this work we take advantage of the cryo-multiplexer platform to advance strained Si/SiGe heterostructures deposited on 300 mm Si substrates in an industrial manufacturing CMOS fab.\cite{pillarisetty2019qubit} The heterostructure comprises a Si$_{0.7}$Ge$_{0.3}$ strained relaxed buffer obtained by step grading of the germanium content, a 10-nm-thick strainded Si quantum well, a 30-nm-thick Si$_{0.7}$Ge$_{0.3}$ barrier and a 1-nm-thick Si cap. Heterostructure-FETs are fabricated in an academic clean room on 100 mm wafers laser-cut from the original 300 mm industrial wafer. In short, the fabrication process for H-FETs involves: mesa-trench for device isolation; P ion implantation and anneal at $T$ = 750 $^\circ$C for contacting the 2DEG; atomic layer deposition of a 30-nm-thick Al$_2$O$_3$ dielectric layer to isolate the 2DEG from the Hall-bar shaped metallic top-gate; metallization for gate, ohmic contacts, and bonding pads. 

\begin{figure}%
	\includegraphics[width=85mm]{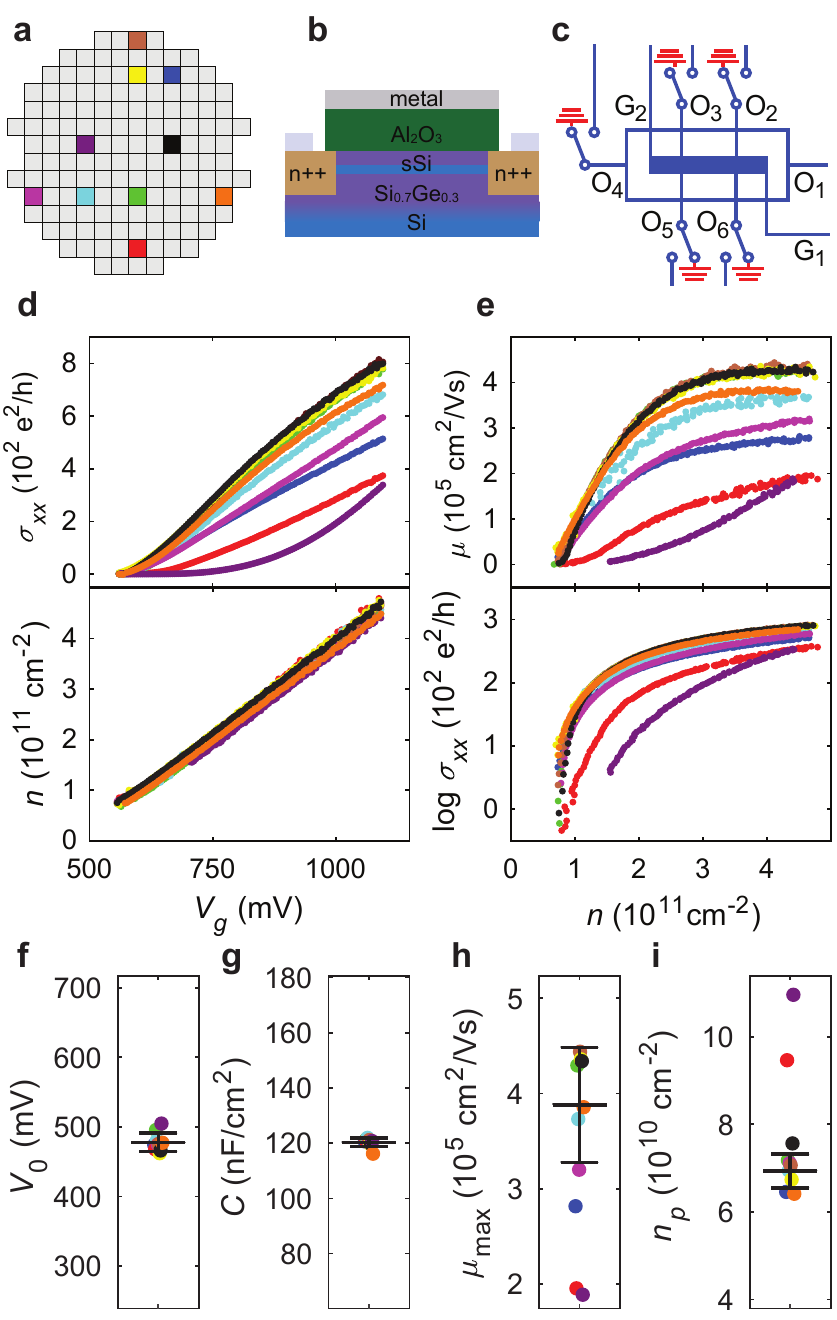}%
	\caption{Multiplexed quantum transport in the classical Hall regime at $T$ = 50 mK. \textbf{a} Dicing scheme of the wafer. Each die and associated measurements throughout the figure have assigned a unique color. \textbf{b} Cross-section schematic of the DUT, a Si/SiGe heterostructure field effect transistor and \textbf{c} contact schematics. \textbf{d} Conductivity $\sigma_{xx}$ (upper panel) and density $n$ (lower panel) as a function of gate voltage $V_g$.  \textbf{e} Mobility $\mu$ (upper panel) and conductivity $\sigma_{xx}$ (lower panel) as a function of density $n$. Statistical analysis of data-sets in \textbf{d} and \textbf{e} yield box plots of \textbf{f} threshold voltage $V_0$, \textbf{g} capacitance $C$, \textbf{h} maximum mobility $\mu_{max}$, and \textbf{i} percolation density $n_p$ with mean and standard deviation (black crosses).  To draw a meaningful comparison between variations in \textbf{f}-\textbf{i} the range of each vertical axis is chosen to equal the mean value of the plotted variable. All measurements are taken in a single cooldown.} 
\label{fig:MAT}
\end{figure}

Ten dies ($N$=10) are randomly selected from different locations of the 100 mm wafer (Fig. \ref{fig:MAT}a), bonded onto the DUT-PCB, and cooled down to 50 mK for measurements.  Fig. \ref{fig:MAT}b and c show a cross-section of the H-FETs and a schematic of the multiple connecting lines, respectively. Each device has 8 terminals. Five ohmic contacts (O2-O6) are multiplexed, whereas the source contact (O1) and the gate contacts (G1, G2) are shared by all $N$ devices ($M$=5, $S$=3). Using these connections we perform magnetotransport measurements on all DUT by standard low frequency lock-in techniques.

We apply a source-drain bias of 0.1 mV and measure $I_{SD}$, the longitudinal voltage $V_{xx}$, and the transverse  Hall voltage $V_{xy}$ as a function of gate voltage $V_G$ and $B$. The longitudinal resistivity $\rho_{xx}$ and transverse Hall resistivity $\rho_{xy}$ are then calculated. The longitudinal ($\sigma_{xx}$) and transverse ($\sigma_{xy}$) conductivity are obtained via tensor inversion. The Hall electron density $n$ is obtained from the linear dependence $\rho_{xy}= B/en$ at low magnetic fields. The carrier mobility $\mu$ is extracted from the relationship $\sigma_{xx}=ne\mu$, where $e$ is the electron charge. 

Figure \ref{fig:MAT}d shows the conductivity and the electron density of the devices measured by time division multiplexing as a function of gate voltage (upper and lower panel, respectively).\footnote{The conductivity measurements are carried out by sequential selection or TDM. As for the control resistance measurements discussed previously, the data-sets agree within less than 1\%, highlighting once more that TDM doesn't perturb the measurements.}  Above a threshold voltage $V_0$, electrons accumulate in the quantum well, current flows in the transistor channel and $\sigma_{xx}$ increases monotonically with $V_0$. Correspondingly, in all devices, the electron density increases linearly as $V_G$ sweeps more positive, consistent with a  parallel‐plate capacitor model where dielectric between the 2DEG and metallic top-gate comprises the Si$_{0.7}$Ge$_{0.3}$ barrier and the Al$_2$O$_3$ layer.

Figure \ref{fig:MAT}e shows the density-dependent mobility and conductivity (upper and lower panel respectively). Excluding the purple and red curves, all the other devices follow a similar trend. The mobility increases steeply at small densities ($n\leq$ 1.4$\times$10$^{11}$cm$^{-2}$), before slowing down and eventually saturating at higher densities ($n\geq$ 2$\times$10$^{11}$cm$^{-2}$). This behaviour is indicative of a high quality Si/SiGe 2DEG. The mobility is limited at low density by scattering from remote charged impurities, likely at the oxide interface, whereas at higher density saturation is given by short-range scattering from impurities within or nearby the quantum well. Remarkably, four devices (black, green, yellow, brown) stand out for exhibiting overlapping mobility density curves over the entire density range, indicating a uniform disorder landscape across the wafer.\cite{sarma2014mobility} This is beneficial for future development of large Si qubit arrays with shared control lines.\cite{li2018crossbar}

By analyzing the data sets in Fig. \ref{fig:MAT}d,e we perform statistical analysis of key metrics of the 2DEG. Threshold voltage, capacitance ($C$), maximum mobility ($\mu_{max}$), and percolation density ($n_p$)are reported as box plots in Fig. \ref{fig:MAT}f-i. The threshold voltage $V_{0}$ (Fig. \ref{fig:MAT}f) is extrapolated from the linear density-gate voltage dependence to zero density, whereas the capacitance (Fig. \ref{fig:MAT}g) is given by the relationship $C = \frac{dn}{dV_g}e$. We observe small variations in both $V_{0}$ (2.75\%) and $C$ (1.34\%) indicating that the dielectric stack comprising a 30-nm-thick Si$_{0.7}$Ge$_{0.3}$ barrier and the Al$_2$O$_3$ layer are uniform across the wafer. A record high $\mu_{max}$ (Fig. \ref{fig:MAT}h) of 4.2$\times 10^5$ cm$^2$/VS is achieved for these industrially manufactured Si 2DEGs, with an average value of (3.9$\pm$ 0.6)$\times10^5$ cm$^2$\slash Vs, corresponding to a standard deviation below 20\%. As expected from the density-dependent mobility curves, the box plot of $\mu_{max}$ reveals the outliers (purple and red), with values outside of the standard deviation. The percolation density $n_p$ (Fig. \ref{fig:MAT}i) is obtained by fitting the density-dependent conductivity to a 2D percolation transition model $\sigma_{xx} \sim (n - n_p)^{1.31}$.\cite{tracy2009observation} We obtain an average $n_p$ of  (6.9$\pm$0.4)$\times10^{10}$ cm$^{-2}$,  corresponding to a standard deviation below 6\%. The percolation density has a minimum value of 6.4$\times 10^{10}$ cm$^{-2}$, on par with the best values reported in the literature.\cite{mi2017circuit,mi2015magnetotransport} Overall these results advance 300 mm epitaxial wafer technology and support the use of wafer-scale Si/SiGe as a promising material platform to manufacture industrial spin qubits.

\begin{figure}%
	\includegraphics[width=85mm]{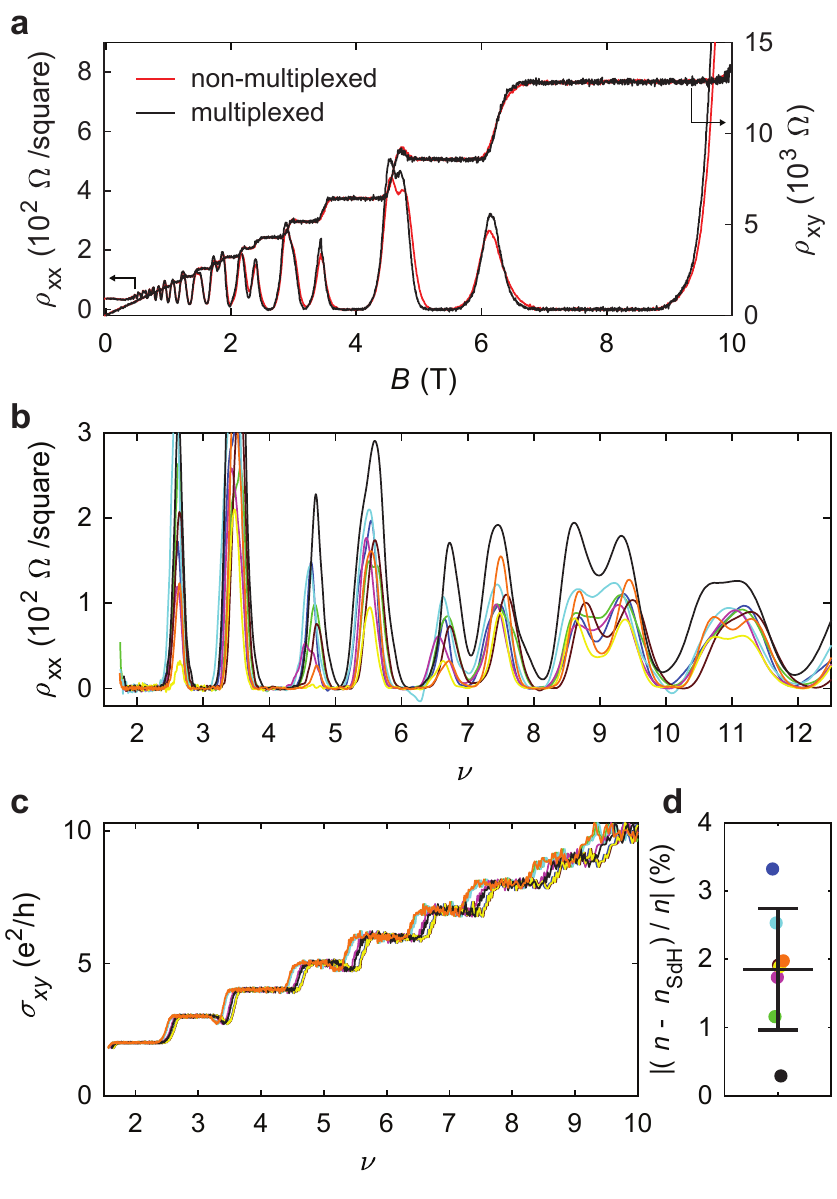}%
	\caption{Multiplexed quantum transport in the classical Hall regime at $T$ = 50 mK. \textbf{a} Resistivity $\rho_{xx}$ and Hall resistivity $\rho_{xy}$ of a Si/SiGe heterostructure field effect transistor at $T$ 50 mK measured individually (red curves) or by multiplexing (black curves) through all other devices. \textbf{b} Multiplexed resistivity $\rho_{xx}$ and \textbf{c} Hall conductivity $\sigma_{xy}$ as a function of filling factor $\nu$ for eight devices. Color coding as described in Fig. \ref{fig:MAT}\textbf{a}. \textbf{d} Box plot of the percentage difference between Hall density $n$ and density $n_{SdH}$, with average and standard deviation (cross), extracted by analysis of the Shubnikov de-Haas oscillation periodicity.} 
\label{fig:SdH}
\end{figure}

We now examine magnetotransport at high magnetic field, where quantum effects are dominant. Figure \ref{fig:SdH}a shows $\rho_{xx}$ and $\rho_{xy}$ of the black device measured either in multiplexed or non-multiplexed conditions. The overlap between the two curves is excellent, confirming the robustness of the setup against magnetic field sweeps.  Clear Shubnikov–de Haas oscillations with zero-resistivity minima are observed in the longitudinal
resistance $\rho_{xx}$ as a function of the magnetic field $B$. Correspondingly, flat quantum Hall effect plateaus are visible in $\rho_{xy}$. The oscillations structure is typical of a Si/SiGe structure. The first oscillations at low fields correspond to integer filling factors $\nu$ = 4$k$ due to the spin and valley degeneracy. At higher fields, opening of the Zeeman gap and increased valley splitting leads to lifting of spin and valley degeneracy and observation of the associated even ($\nu$ = 4$k$-2) and odd ($\nu$ = 2$k$-1) filling factors. The QHE plateaus values are quantized as expected at values of $h/e^2v$, where $h$ is Planck’s constant and $e$ the elementary charge.

Figure \ref{fig:SdH}b shows the multiplexed $\rho_{xx}$ measurements for all devices excluding the purple and red device.\footnote{Analysis of the Shubnikov -de Haas oscillations reveals for the purple and red device the presence of a spurious conduction channel in parallel to the quantum well, possibly due to leakage from the gate. This spurious channel is likely cause of the reduced mobility compared to the other devices at similar density.} The measurement are taken at a fixed $V_G$, corresponding to $n\sim$4.3$\times$10$^{11}$ cm$^{-2}$. For clarity, the curves are plotted against filling factor $\nu$. All devices show clearly the spin and valley split levels, however differences in the values of $\rho_{xx}$ are seen, possibly due to the different Landau level broadening and/or different energy splittings across devices. Similar considerations apply to the minor difference observed in quantum Hall measurements reported in Fig. \ref{fig:SdH}c. As a final statistical analysis, we show in Fig. \ref{fig:SdH}d a box plot of the percentage difference between Hall density and Shubnikov -de Haas density $n_{SdH}$, obtained by the periodicity of the oscillations as a function of $1/B$. The discrepancy is less than 3\%, indicating that population of only one high-mobility subband is achieved uniformly across the wafer.

\section{Discussion}

In conclusion, we investigate a cryo-CMOS architecture that uses low-cost discrete components at 50 mK to increase the number of wires available at cryogenic temperature by an order of magnitude. This is obtained while keeping the overhead number of I/O wires at room temperature fixed. As a proof of principle, we develop and operate a cryo-MUX PCB with 16 selectable channels and 6 multiplexed lines, resulting in 96 wires available at the base temperature of the dilution refrigerator. This solution, implemented in a dilution refrigerator insert with a small sample space, can be further expanded and readily applied to virtually any cryostat. 

We show control experiments where time-division multiplexed measurements of known resistors are performed to demonstrate robustness of the setup with respect to applied voltages, magnetic field, and temperature sweeps. We harness the power of the multiplexing architecture to measure quantum transport of numerous Si/SiGe H-FETs in one cooldown, advancing 300 mm Si/SiGe wafers fabricated in an industrial CMOS fab to record values of maximum mobility and percolation density.  Further improvements of these two metrics are expected by processing the entire gate stack in the high volume manufacturing environment, due to the better semiconductor/oxide interface attainable with an advanced process control. Multiplexed measurements of Shubnikov de-Haas oscillations and quantum Hall effect are performed successfully. These capabilities provide scope for future high-volume measurements of valley splitting in Si 2DEGs based on thermal activation measurements in the QHE regime. 

We show a path forward for high-throughput quantum transport at cryogenic temperatures which will help to accelerate the fabrication-measurement cycle of quantum devices in industrial settings. Furthermore, we consider our setup as a blueprint for more complicated electrical architectures, such as 2D-arrays for spin-qubits,\cite{li2018crossbar,veldhorst2017silicon,vandersypen2017interfacing} since a 100\% switching fidelity is achieved and the multiplexers support switching in the the MHz regime. We envisage that investigations of different components with smaller footprints, circuits, and architectures at cryogenic temperatures, including custom fully-integrated CMOS solutions, will help to satisfy the ever growing need for scalable wiring solutions to control large quantum systems.

\section*{Acknowledgements}
This work is part of the research programme OTP with project number 16278, which is (partly) financed by the Netherlands Organisation for Scientific Research (NWO). M.V. acknowledges support from the NWO Vidi Program. Research was sponsored by the Army Research Office (ARO) and was accomplished under Grant No. W911NF- 17-1-0274. The views and conclusions contained in this document are those of the authors and should not be interpreted as representing the official policies, either expressed or implied, of the Army Research Office (ARO), or the U.S. Government. The U.S. Government is authorized to reproduce and distribute reprints for Government purposes notwithstanding any copyright notation herein. The authors would like to thank N. Alberts mechanical engineering support and K. Loh and M. W. L. Wilmer for software support.

\end{document}